\documentclass[12pt]{article}
\usepackage[pctex32]{graphics}
\begin{document}
\vspace{1cm}
\begin{center}
~
\\
~
{\bf  \Large Statistics of Anyon Gas and the Factorizable Property of Thermodynamic Quantities}
\vspace{1cm}

                      Wung-Hong Huang\\
                       Department of Physics\\
                       National Cheng Kung University\\
                       Tainan, Taiwan\\

\end{center}
\vspace{1cm}
The statistical distribution function of anyon is used to find the eighth viral coefficient in the high-temperature limit and the equation of state in the low-temperature limit. The perturbative results indicate that the thermodynamic quantities, $Q(\alpha)$, of the free anyon gas may be factorized in the terms characteristic of the ideal Bose ($\alpha =0$) and fermion ($\alpha =1$) gases, i.e., $Q(\alpha) = \alpha Q(1) + (1-\alpha) Q(0)$. It is shown that the factorizable property of the thermodynamic quantities, to all orders, can be established from the property of the equivalence between the anyon statistics and statistics in a system with boson-fermion transmutation, which was found by us in a recent paper (hep-th/0308095; Phys. Rev. E 51 (1995) 3729)
\vspace{2cm}
\begin{flushleft}
Phys. Rev B53 (1996) 15842
\\
*E-mail:  whhwung@mail.ncku.edu.tw\\
\end{flushleft}
\newpage
\section{Introduction}
Anyons are the quantum objects which obey fractional  statistics and exist in two dimensions [1,2]. In recent years, it was realized that anyons may be regarded as the quasiparticles in the fractional quantized Hall effect [3,4]. It has also been proposed that anyons may play an important role in high-temperature superconducting [5-7]. The thermodynamic behavior of anyon gas has been studied by several techniques, including a perturbative expansion about fermion6 [6-8] or boson [9,10]. The viral coefficient of free anyon has also been calculated [11,12]. 
 
Most studies have been done in the context of many-body quantum mechanics. In recent papers [13,14], Wu and others had derived the occupation-number distribution function of the anyon gas to formulate the theory of quantum statistical mechanics, with the help of the idea of fractional exclusion statistics [15]. In that paper [13], Wu also obtained the second viral coefficient and found that the ``statistical interaction"  may be attractive or repulsive, depending on the statistical parameter.

 In this paper, we will use the statistical distribution function of anyon to evaluate the eighth viral coefficient in the high-temperature limit and the equation of state in the low-temperature limit. From the calculated results we see that, to the order of perturbation, the thermodynamic quantities of  $N$ anyon gas, denoted as $Q_N(\alpha)$, can be factorized in the terms characteristic of the ideal Bose ( $\alpha =0$) and fermion ($\alpha =1$) gases, i.e., $Q_N(\alpha) = \alpha  Q_N(1) + (1-\alpha) Q_N(0)$. Furthermore, with the help of the ``boson-fermion transmutation interpretation" of anyon statistics, which was found by us in a recent paper [14], it is interesting to see that the factorizable property can be established to all orders of perturbation.

\section {High-Temperature and Low-Temperature Expansion}
Using the Haldane's fraction exclusion statistics [15], the 
statistical distribution function derived in Refs. 13 and 14 is 
$$n(\epsilon) = {1\over W(\zeta)+\alpha},\eqno{(2.1)}$$ 
where 
$$W(\zeta)^\alpha [1+W(\zeta)]^{(1-\alpha)}=\zeta,~~~\zeta = e^{(\epsilon-\mu)/kT},\eqno{(2.2)}$$ 
and $\alpha$ is the fraction statistical parameter. The above relations 
recover the familiar Bose and fermion distributions, respectively, with $\alpha=0$ and $\alpha=1$.   Wu [13] had also derived the relations 
$${\mu\over kT}= \alpha {2\pi \hbar^2\over m kT}{N\over V}+ ln \left[1-exp\left(-{{2\pi \hbar^2\over m kT}{N\over V}}\right)\right],\eqno{(2.3)}$$ 
$$PV = E = {V m\over 2\pi \hbar^2}\int_0^\infty d\epsilon \epsilon n(\epsilon).\eqno{(2.4)}$$ 
We now use the above relations to perform the high-temperature and low-temperature expansion of equation of state through the standard manipulations [16].  

\subsection{A. High-Temperature Expansion} 
In the Boltzmann limit $\zeta >> 1$, and thus $W(\zeta) >>1$, we can use (2.2) to find the high-temperature expansion of function $W(\zeta)$. Then, substituting the $W(\zeta)$ into (2.1), we have 
$$ n(\zeta) ={1\over\zeta} +(1-2\alpha){1\over\zeta^2} +{2-9\alpha+9\alpha^2\over2\zeta^3}+{3-22\alpha+48\alpha^2-32\alpha^3\over3\zeta^4}$$
$$+{24-250\alpha+875\alpha^2-1250\alpha^3+625\alpha^4\over24\zeta^5}$$
$$+{10-137\alpha+675\alpha^2-1530\alpha^3+1620\alpha^4-648\alpha^5\over10\zeta^6}+...\eqno{(2.5)}$$
Note that we only present the perturbative form of $n(\epsilon)$ to the order of $\zeta^{-6}$ in the above relation; however, the calculation presented in the following is the result of the perturbation to the order of $\zeta^{-9}$. Substituting the high-temperature expansion of $exp(\mu/kT)$, with the help of (2.3), into (2.5) and performing the integration of $\epsilon$ (2.4), the viral expansion of the equation of state becomes
$$PV = NkT\left[1+\sum_\ell^\infty b_{\ell +1}(\alpha)\left({N\over V}\lambda^2\right)^\ell~\right]= NkT\left[1+{1\over2}\left(\alpha-{1\over2}\right){N\over V}\lambda^2~+{1\over36}\left({N\over V}\lambda^2\right)^2\right.$$
$$\left.  -{1\over3600}\left({N\over V}\lambda^2\right)^4+ {1\over211680}\left({N\over V}\lambda^2\right)^6+ O\left(\left({N\over V}\lambda^2\right)^8\right)\right],\eqno{(2.6)}$$
where the thermal wavelength $lambda = (2\pi\hbar^2/mkT)^{1/2}$ Note that Sen[12] had calculated the viral coefficient $b_4$ in the context of many-body quantum mechanics. He suspects, but has not completely proven, that all the higher-order viral coefficient $b_l(\alpha) = b_l(0)+O(\alpha^2)$ for all $\ell >4$. On the other hand, we have calculated the viral coefficient $b_7$ in (2.7) in 
the context of quantum statistical mechanics and it is consistent with Sen's conjecture. Note also that, recently, Murthy and Shanker [17] have shown that in the Haldane fraction statistics, the fraction statistical parameter is completely determined by the high-temperature limit of the second viral coefficient. 
Our result is consistent with their conclusion. 

We have noticed that in the two-dimensional space the energy of boson gas or fermion gas is given by [13] 
$$E = NkT\sum_\ell^\infty {B_{\ell}\over (\ell +1)!}\left({N\over V}\lambda^2\right)^\ell~~~~~~~~(boson~gas),\eqno{(2.7a)}$$
$$E = NkT\sum_\ell^\infty (-1)^\ell{B_{\ell}\over (\ell +1)!}\left({N\over V}\lambda^2\right)^\ell~~~~~~~~(fermion~gas),\eqno{(2.7b)}$$
where $B_\ell$ are the Bernoulli number with $B_0=1$, $B_1=-{1\over2}$, $B_2={1\over6}$, and $B_{2\ell+1}=0$, for $\ell \ge1$. This indicates that, to the 
order that we have calculated, there has been an interestingly factorizable property, 
$$E(\alpha) = \alpha E(1) + (1-\alpha)E(0).\eqno{(2.8)}$$
We will, in the next section, show that the factorizable property of the thermodynamic quantity, to all orders, can be easily established from the property of the equivalence between the anyon statistics and statistics in a system with 
boson-fermion transmutation. (The equivalent property was proven in our previous paper [14]. As the factorizable property does not depend on the temperature, it may thus also be found in the low-temperature expansion of the equation of state, which will be calculated in Sec. II B.  Note also that 
because the viral expansion for the free boson and fermion gases only differ in the second coefficient, as expressed in (2.6), the factorization formula thus leads to a property that only the second viral coefficient can depend on the fraction 
statistical parameter $\alpha$, the result shown by Murthy and Shanker [17]. 

\subsection{B. Low-Temperature Expansion} 
For the boson gas, $\alpha=0$, we have 
$$PV = {Vm\over2\pi\hbar^2}\int_0^\infty d\epsilon \epsilon {1\over e^{(\epsilon-\mu)/kT}-1}= {Vm\over2\pi\hbar^2}(kT)^2 \int_0^\infty dx x {1\over \left[{1\over 1-exp(-\lambda^2N/V)}\right]e^x-1}$$
$$ = {Vm\over2\pi\hbar^2}(kT)^2\left[{\pi^2\over6}+O(e^{\lambda^2N/V})\right].\eqno{(2.9)}$$
If $\alpha \ne 0$, then through the standard procedure [16] we have 
$$PV = {Vm\over2\pi\hbar^2}\int_0^\infty d\epsilon \epsilon ~{1\over {W[e^{(\epsilon-\mu)/kT}] +\alpha}}\hspace{9cm}$$
$$={Vm\over2\pi\hbar^2}\left[\int_0^\mu d\epsilon \epsilon \left[{1\over\alpha} - {1\over {\alpha +\alpha^2W[e^{(\epsilon-\mu)/kT}]}}\right]+ \int_\mu^\infty d\epsilon \epsilon ~{1\over {W[e^{(\epsilon-\mu)/kT}] +\alpha}} \right]$$
$$ = {V2\pi \hbar^2\over m}\left({\alpha\over2}\right)\left({N\over V}\right)^2+ {Vm\mu\over2\pi\hbar^2} (kT)\int_0^\infty dx \left[{1\over W(e^x)+\alpha} - {1\over\alpha+\alpha^2 W(e^x)}\right] $$
$$+{Vm\mu\over2\pi\hbar^2}(kT)^2\int_0^\infty dx x\left[{1\over W(e^x)+\alpha} + {1\over\alpha+\alpha^2 W(e^x)}\right].\eqno{(2.10)}$$
The integration of the second term in the above equation can  be shown to be zero by the following observation: 
$$N =  {Vm\over2\pi\hbar^2}\int_0^\infty d\epsilon ~{1\over {W[e^{(\epsilon-\mu)/kT}] +\alpha}}\hspace{6cm}$$
$$ ={Vm\over2\pi\hbar^2} {\mu\over\alpha}+ {Vm\over2\pi\hbar^2}(kT) \int_0^\infty dx \left[{1\over W(e^x)+\alpha} - {1\over\alpha+\alpha^2 W(e^x)}\right] .\eqno{(2.11)}$$
Comparing the above relation with the low-temperature limit of  (2.3), we see that the integration in the above equation is zero. 

It is unfortunate that we do not yet have an ability to show that the integration in the third term of  (2.10) does not depend on the fractional statistical parameter $\alpha$. However, for the special cases including $\alpha=1$, ${1\over2}$, ${1\over3}$, ${1\over4}$, and cases with small $\alpha$  the analytic function of  $W(e^x)$ can be obtained and  value of integration is $\pi^2/6$ for these cases. Combining the results of $\alpha\ne 0$ with the boson case, $\alpha =0$, we see that the factorization formula  (2.8) can also be used in these cases. 

\section{Factorization and Boson-Fermion Transmutation}
In this section, we will show that the thermodynamic quantity, such as energy, heat capacity, and entropy, etc., can be expressed in the factorized form. The derivation is very simple and is only based on the property of the equivalence 
between the anyon statistics and statistics in a system with boson-fermion transmutation. 

In our recent paper [14], we have shown that, for a system with an $\alpha$ fraction of fermion and a ($1-\alpha$) fraction of boson, once the transmutation between the boson and fermion is allowed, the system will have a statistical distribution function just like that of free anyon. 

It is easy to see that the system with boson-fermion transmutation can also be regarded as the ensemble average of the $M$ systems, which are classified as the fermion case and the boson case. In the fermion case, there are $\alpha M$ systems and each one has $N$ fermion gas existing in the volume $V$ and has 
the pressure $P$. In the boson case, there are $(1-\alpha)M$ systems and each one has $N$ boson gas existing in the volume $V$ and has the pressure $P$. Therefore, the ensemble average for the thermodynamic quantity $Q_N(\alpha)$ of the system with the boson-fermion transmutation, and thus the anyon system, can be factorized as 
$$ Q_N(\alpha) = \alpha Q_N(1) + (1-\alpha)Q_N(0). \eqno{(3.1)}$$
This completes our proof. 

It has many simplifications during the investigation of the thermodynamic behavior of anyon gas, once the factorized formula has been set up. The reason for this is that if we try to calculate the thermodynamic quantities of anyon from the anyon distribution function, then the nonanalytic function $W(\zeta)$ in  (2.2), which can only be solved analytically in the general case, will make it difficult to study. On the other hand, if we use the factorized formula to calculate the thermodynamic behaviors of anyon, then the known function $W(\zeta)$ in the boson and fermion cases will make it easy to study. 

Finally, it shall be mentioned that Murthy and Shankar [18] had also found the factorizable property in an anyon system They investigated the Calogero-Sutherland model and found that this model obeys the statistical distribution function derived by Wu [13,14]. The factorizability of the anyon partition 
they found is crucially based on the special energy spectrum in the Caloger-Sutherland model. On the other hand, in our proof, we do not rely on any property of the anyon system and this means that the factorizable property of the anyon gas is a very general property. 
~
\\
~
\\
~
\\
~
{\bf  \Large References}
\begin{enumerate}
\item  J. M. Leinaas and J. Myrheim, Nuovo Cimento B 38, 1 (1977). 
\item  F. Wilczek, Phys. Rev. Lett. 48, 1144  (1982); 49, 957  (1982). 
\item R. B. Laughlin, Phys. Rev. Lett. 50, 1359 (1983); Phys. Rev. B27, 3383 (1983!; B. I. Halperin, Phys. Rev. Lett. 52, 1583 (1984); 52, 2390 E!(1984); D. Arovas, J. R. Schrieffer, and F. Wilczek, ibid. 53, 722 (1984). 
\item R. Prange and S. M. Girvin, The Quantum Hall Effect ~Springer, 
Berlin, (1987). 
\item  P. B. Wiegmann, Phys. Rev. Lett. 60, 821 ~1988!; R. B. Laughlin, 
ibid. 60, 2677 (1988). 
\item A. L. Fetter, C. B. Hanna, and R. B. Laughlin, Phys. Rev. B 39, 9679  (1989); C. B. Hanna, R. B. Laughlin, and A. L. Ferrer, ibid. 40, 8745 (1989). 
\item Y.-H. Chen, F. Wilczek, E. Witten, and B. I. Halperin, Int. J. Mod. Phys. B 3, 1001 (1989); B. I. Halperin, J. March-Russell, and F. Wilczek, Phys. Rev. B 40, 8726  (1989). 
\item Y. Hosotani and S. Chakravarty, Phys. Rev. B 42, 342 (1990). 
\item X. G. Wen and A. Zee, Phys. Rev. B 41, 240 (1990); H. Mori, ibid. 42, 184 (1990). 
\item D. Sen and R. Chitra, Phys. Rev. B 45, 881  (1992). 
\item D. Arovas, J. R. Schrieffer, F. Wilczek, and A. Zee, Nucl. Phys. B 251, 117 (1985). 
\item D. Sen, Nucl. Phys. B 360, 397  (1991). 
\item  Y.-S. Wu, Phys. Rev. Lett. 73, 922  (1994); C. Nayak and F. Wilczek, ibid. 73, 2740 (1994); A. K. Rajagopal, ibid. 74, 1048 (1995); Wung-Hong Huang, ``Statistical Interparticle Potential between Two Anyons," Phys. Rev. B52 (1995) 15090 [hep-th/0701715].
\item Wung-Hong Huang, ``Boson-Fermion Transmutation and Statistics of Anyon," Phys. Rev. E51 (1995) 3729  [hep-th/0308095].
\item  F. D. M. Haldane, Phys. Rev. Lett. 67, 937 (1991); M. D. Johnson 
and G. S. Canright, Phys. Rev. B 49, 2947 (1994). 
\item R. K. Pathria, Statistical Mechanics ~Pergamon, London, (1972); 
K. Huang, Statistical Mechanics ~Wiley, New York, (1963). 
\item M. V. N. Murthy and R. Shankar, Phys. Rev. Lett. 72, 3629 (1994). 
\item M. V. N. Murthy and R. Shankar, Phys. Rev. Lett. 73, 3331 (1994).
\end{enumerate}
\end{document}